\def\be{\begin{equation}}
\def\te{\end{equation}}
\def\bea{\begin{eqnarray}}
\def\nn{\nonumber}
\def\tea{\end{eqnarray}}

\def\a{\alpha}
\def\b{\beta}

\def\d{\delta}
\def\e{\epsilon}

\def\g{\raisebox{.4ex}{$\gamma$}}

\def\o{\omega}
\def\p{\pi}
\def\q{\theta}

\def\t{\tau}

\def\z{\zeta}
\def\D{\Delta}

\def\G{\Gamma}

\newskip\humongous \humongous=0pt plus 1000pt minus 1000pt

\newif\ifdtup

\def\a{\alpha}

\def\ha{{1\over 2}}

\def\tz{t_0}
\def\xz{x_0}
\def\uz{u_0}
\def\vz{v_0}
\def\tah{\tau_H}
\def\taz{\tau_0}
\def\intf{\int_{0}^{\infty}}
\def\intin{\int_{-\infty}^{\infty}}

\newcommand{\tbar}{\overline t}

\newcommand{\eqn}[1]{(\ref{#1})}

\documentstyle[12pt]{article}

\makeatletter                    
\@addtoreset{equation}{section}  
\makeatother                     

\begin{document}

\title{Near-Thermal Radiation in Detectors, Mirrors
and Black Holes: A Stochastic Approach}
\author{Alpan Raval\thanks{E-mail: raval@umdhep.umd.edu},
B. L. Hu\thanks{E-mail: hu@umdhep.umd.edu}\\
{\small Department of Physics, University of Maryland,
College Park, MD 20742, USA}\\
Don Koks\thanks{E-mail: dkoks@physics.adelaide.edu.au} \\
{\small Department of Physics, University of Adelaide,
Adelaide 5005, Australia}}
\date{{\small \it (gr-qc/9606074, umdpp 96-53. Submitted to
Phys. Rev. D
on June 26, 1996)}\\
{\small \it PACS number(s):  04.62.+v, 32.80.-t,
05.40.+j, 42.50.Lc, -04.70.-s.}}
 \maketitle

\begin{abstract}
In analyzing the nature of thermal radiance experienced by 
an accelerated observer (Unruh effect), an eternal black hole
(Hawking effect)
and in certain types of cosmological expansion,
one of us proposed a unifying viewpoint that these can be
understood as
arising from the vacuum fluctuations of the quantum  field
being subjected to an exponential scale transformation in these
systems.
This viewpoint, together with our recently developed stochastic
theory of particle-field interaction understood as quantum open
systems
described by the influence functional formalism,
can be used effectively to address situations where the spacetime
possesses an event horizon only asymptotically, or none at all.
Examples studied here include detectors moving
at uniform acceleration only asymptotically or for a finite time,
a moving mirror, and a
two-dimensional collapsing mass. We show that in such systems
 radiance indeed is
observed, albeit not in a precise Planckian spectrum.
The deviation therefrom is determined by a parameter which
measures the departure from  uniform acceleration or from
exact exponential expansion.
These results are expected to be useful for the investigation of 
non-equilibrium black hole thermodynamics and the linear response 
regime of backreaction problems in semiclassical gravity.

\end{abstract}

\newpage
\section{Introduction}

Particle production \cite{cospc} with a thermal spectrum
from black holes \cite{Bek73,Haw75,bhpc}, moving mirrors
\cite{FulDav},
accelerated detectors \cite{Unr76}, observers
in  de Sitter Universe \cite{GibHaw} and certain cosmological
spacetimes \cite{cosrad} has been a subject of continual discussion
since the mid-seventies because of its extraordinary nature and its basic
theoretical value.
The mainstream approach to these problems relied on thermodyamic
arguments \cite{DavTDBH,WaldTDBH},
finite temperature field theory techniques
\cite{bhprog,DowCriKen,tftcos},
or geometric constructions
(event horizon as a global property of spacetime) \cite{York},
or pairwise combinations thereof. The status of work
on quantum field theory in curved spacetimes up to 1980 can be found
in \cite{BirDav}.
The eighties saw attempts and preparations for the backreaction
problem \cite{bhbkrn80}
(for cosmological backreaction problems, see \cite{cpcbkr}), i.e.,
the calculation of the energy momentum tensor (see \cite{And}
and earlier references),  the effect of particle creation on a
black hole
(in a box, to ensure quasi-equilibrium with its radiation)
\cite{York}, and the dynamical origin of black hole entropy
\cite{FroNov}.
These inquiries are mainly confined to equilibrium thermodynamics
or finite-temperature field theory conditions.%
\footnote{Among other notable alternatives, we'd like to
mention Sciama's dissipative system approach \cite{Sciama},
Unruh's work on sonic black holes \cite{UnrSonBH}
(see also Jacobson \cite{JacSonBH}), Zurek and Thorne's degree of
freedom counts \cite{ZurTho},
Sorkin's geometric or `entanglement' entropy \cite{Sor,GeomEnt}
(see also \cite{HuKan}), and Bekenstein-Page's information theory
approach \cite{BekInfEnt,Page}).
See also the views expressed earlier by Stephens, t'Hooft and
Whiting \cite{tHo}.}
To treat problems of a dynamical nature such as the backreaction
of Hawking radiation on black hole collapse, one needs a new
conceptual framework and a
more powerful formalism for tackling non-equilibrium conditions
and high energy (trans-Planckian) processes.
A new viewpoint which stresses the local, kinematic nature of
these processes rather than the traditional global geometric
properties  has been proposed \cite{HuEdmonton,cgea,jr,Dalian}
which regards the Hawking-Unruh thermal radiance observed in
one vacuum as resulting from  exponential redshifting of
quantum noise of another. This view puts the nature of thermal
radiance in the two classes of
spacetimes on the same footing \cite{therad},
and enpowers one to tackle situations which do not possess an event
horizon at all,  as the examples in this and a companian paper
will show.

Such a formalism of statistical field
theory has been developed by one of us and co-workers in recent
years \cite{CalHu,HuPhysica,HuTsukuba,HuWaseda,Banff}.
This approach aims to provide the quantum statistical
underpinnings of field theory in curved spacetime, and strives at a
microscopic and elemental description of the structure and
dynamics of matter  and spacetime. The starting point is the
quantum and thermal fluctuations for fields, the focus is on the
evolution of the reduced density matrix of an open system
(or the equivalent distribution or Wigner functions);
the quantities of interest are the noise and dissipation kernels
contained in the influence functional \cite{if},
and the equation of motion takes the form of
a master, Langevin, Fokker-Planck, or stochastic Schr\"odinger
equation describing the evolution of the quantum statistical
state of the system,
including, in addition to the quantum field effects like
radiative corrections and renormalization, also statistical
dynamical effects like decoherence, correlation  and dissipation.
Since it contains the causal (Schwinger-Keldysh) effective action
\cite{ctp} it is a generalization of the traditional scheme of
thermal field theory \cite{tftcos}
and the `in-out' (Schwinger-DeWitt) effective action \cite{SchDeW},
and is particularly suited for treating fluctuations and dissipation
in backreaction problems in semiclassical gravity \cite{Banff}.

The foundation of this approach has been constructed recently
based on the open system concepts and the
quantum Brownian model \cite{HPZ,HM2}. The method has since been
applied to particle creation and backreaction processes in
cosmological spacetimes
\cite{nfsg,HM3,fdrsc}. For particle creation in spacetimes with
event horizons, such as for an accelerated observer and black holes,
this method derives
the Hawking and Unruh effect \cite{Ang,HM2}
from the viewpoint of exponential amplification of quantum noise
\cite{Dalian}. It can also describe the linear response regime of
backreaction viewed as a fluctuation-dissipation relation.
\cite{fdrsc,RHA}.

This paper is a continuation of our earlier work \cite{HM2,Ang,RHA}
to present two main points:\\
1) {\it A unified approach} to treat thermal particle creation from
both spacetimes with and without event horizons \cite{therad}
based on the interpretation that the thermal radiance can be viewed
as resulting from quantum noise
of the field being amplified by an exponential scale transformation
in these systems (in specific vacuum states) \cite{Dalian}.
In contradistinction to viewing these as global, geometric effects,
this viewpoint emphasizes the kinematic effect of scaling on
the vacuum in altering the relative weight  of quantum versus
thermal fluctuations.\\
2){\it An approximation scheme} to show that near-thermal radiation
is emitted from systems undergoing near-uniform acceleration or in
slightly perturbed spacetimes. We wish to demonstrate the relative
ease in constructing
perturbation theory using the statistical field theory methods.
Let us elaborate on these two points somewhat.

It may appear that this approximation can be equally implemented
by taking the
conventional viewpoints (notably the geometric viewpoint),
and the perturbative calculation can be performed by other existing
methods (notably the thermal field theory method). But as we will
show here, it is not as easy as it appears.
{\it Conceptually}, the geometric viewpoint assumes that a sufficient
condition for the appearance of Hawking radiation is the existence
of an event horizon, which is considered as a global property of the
spacetime or the system.
(Note for the case of an extreme Reissner-Nordstrom spacetime, this
is not the case, as there exists an event horizon but no radiation.
\footnote{This is the one case which arose in the discussion between
Hu and Unruh (private communication, 1988) who shared the somewhat
unconventional view that the exponential redshifting is a more basic
mechanism than event horizons responsible for thermal radiance.})
When the spacetime deviates from the eternal black hole, or that the
trajectory deviates from the uniformly accelerated one, physical
reasoning tells us that the Hawking or Unruh radiation should still
exist, albeit with a non-thermal spectrum.
But the event horizon, if exists,  of the deformed spacetime
may not be so easily described in geometric terms. And for
time-dependent perturbations of lesser symmetry, or for situations
where uniform acceleration occurs only for a finite interval of time,
it is not easy to deduce the
form of Hawking radiation in terms of purely global geometric
quantities (see, however, Wald \cite{Wald} and Teitelboim et al
\cite{BTZ}).
The concept of an approximate event horizon,
which exists for a finite period of time or only asymptotically, is
difficult to define and even if it is possible (by apparent
horizons, e.g., \cite{YorkAppHor}), rather unwieldy to implement in
the calculation of particle creation and backreaction effects.

{\it Technically} one may think calculations via the thermal field
theory are equally possible. Indeed this has been tried before by
one of us and others.
One way is to assume a quasi-periodic condition on the Green function,
making it near-thermal \cite{tftcos}. But this is not a good solution,
as the deviation from eternal black hole or uniform acceleration
disables the Euclidean section in the spacetimes (Kruskal or Rindler),
and the imaginary time finite temperature theory is not well defined
any more. Besides, to deal with the statistical dynamics of the system,
one should use an in-in boundary condition and work with causal
Green functions. The lesson  we learned in treating the backreaction
problems of
particle creation in cosmological spacetimes \cite{CalHu,nfsg}
is that one can no longer rely on methods which are restricted to
equilibrium conditions (like the imaginary-time or thermo-field
dynamics methods), but should
use nonequilibrium methods such as
the Schwinger-Keldysh (closed time path) effective action
\cite{ctp} for the treatment of backreactions. Its close equivalent,
the influence
functional method \cite{if} is most appropriate for investigating
the statistical mechanical aspects of matter and geometry, like the
entropy of quantum fields and
spacetimes, information flow, coherence loss, etc.~\cite{Erice95}.

In this and a companian paper \cite{KHRM} we shall use these methods
to analyze particle creation in perturbed situations whose background
spacetime possess an event horizon, such as an asymptotically
uniformly-accelerated detector, or one accelerated in a finite time interval
(Sec.\ 3),
the moving mirror (Sec.\ 4) and the asymptotically Schwarzschild
spacetime (Sec.\ 5).
In the follow-up paper we shall study near-thermal particle
creation in an exponentially expanding universe, a slow-roll
inflationary universe, and a universe in asymptotically-exponential
expansion. What ties the problem of thermal radiation in cosmological
as well as black hole spacetimes together is the exponential
scale-transformation
viewpoint expressed  earlier \cite{HuEdmonton,cgea,jr,Dalian,therad}.
The stochastic theory approach is capable of implementing
this view. One can describe all these systems with a single parameter
measuring the deviation
from uniformity or stationarity, and show that the same parameter
also appears in the near-thermal behavior of particle creation in all
these systems. This result will be used later in our exploration of the
linear-response regime of the backreaction problem in semiclassical
gravity.

\subsection{General Formalism}

Consider a particle detector linearly coupled to a quantum field.
The dynamics of the internal coordinate $Q$ of the detector in a
wide class of spacetimes is derived in \cite{HM2}, and can be
described by Langevin equations of the form :
\be
\frac{\partial L}{\partial Q} - \frac{d}{dt}
\frac{\partial L}{\partial \dot{Q}}
-2\int_{t_i}^{t}\mu(t,s)Q(s)ds = \xi (t)
\te
where $\xi(t)$ is a stochastic force with correlator
$\langle\xi(t)\xi(t')\rangle = \hbar \nu(t, t')$. The trajectory
$x^{\mu}(s)$ of the detector, parametrized by a suitable parameter $s$,
will be denoted simply by $x(s)$ for convenience.

For the special case of linear coupling between a field $\phi$ and
the detector of the form $L_{int} = eQ\phi(x(s))$,  the kernels
$\mu$ and $\nu$, called the {\it dissipation} and {\it noise}
kernels respectively, are given by
\bea
\mu(s, s') &=& \frac{e^2}{2}G(x(s),x(s')) \equiv -i\frac{e^2}{2}
<[\hat{\phi}(x(s)), \hat{\phi}(x(s'))]> \\
\nu(s, s') &=& \frac{e^2}{2}G^{(1)}(x(s),x(s')) \equiv \frac{e^2}{2}
<\{\hat{\phi}(x(s)), \hat{\phi}(x(s'))\}>
\tea
where $G$ and $G^{(1)}$ are the Schwinger and the Hadamard functions
of the free field operator $\hat{\phi}$ evaluated for two points
on the detector trajectory, $<\,>$ denotes expectation value with respect
to a vacuum state at some arbitrarily chosen initial time $t_i$,
and $[\,,\,]$ and $\{\,,\,\}$ denote the commutator and
anticommutator respectively. This result may
be obtained either by integrating out the field degrees of freedom
as in the Feynman-Vernon influence functional approach [40] or
via manipulations of the coupled detector-field Heisenberg
equations of motion in the canonical operator approach.

It will often be convenient to express the kernels $\mu$ and $\nu$
as the real and imaginary parts of a complex kernel $\zeta \equiv \nu + i\mu $,
called the {\it influence kernel}. For linear
couplings, it follows from the above expressions that $\zeta $ is
given by the
Wightman function $G^{+}$:
\be
\zeta(s, s') = e^2 G^{+}(x(s), x(s')) \equiv e^2 <\hat{\phi}(x(s))
\hat{\phi}(x(s'))>.
\te
The influence kernel thus admits the mode function representation
\be
\zeta(s, s') = e^2\sum_k u_k(x(s))u_k^{\ast}(x(s')),
\te
the $u_k$'s
being the fundamental modes satisfying the field equations and
defining the particular Fock space whose vacuum state is the one
chosen above. This method of evaluating
the kernels $\mu$ and $\nu$ is only applicable for linear
coupling cases.

An alternative approach [44], consists of decomposing the field
Lagrangian into parametric oscillator Lagrangians at the very outset,
thus converting a quantum field-theoretic problem to a quantum
mechanical one. Denoting the parametric oscillator degrees of
freedom by $q_k$ (and their masses and frequencies by $m_k$ and
$\omega_k$ respectively), the detector-field interaction mentioned
earlier
is generally given by $L_{int} = \sum_k c_k(s) Q q_k$, where the
coupling ``constants'' $c_k$ now become time-dependent, and
contain information about the detector trajectory. In this approach,
the influence kernel is given in terms of the oscillator mode
functions $X_k$, as
\be \label{blue32.4-1}
\zeta(s, s') = \int_0^\infty dk\;I(k,s,s')X_k(s)X_k^\ast(s')
\te
where the $X_k$'s  satisfy the parametric oscillator equations
\be
\ddot{X_k} + \omega_k^2(t) X_k = 0
\te
satisfying the initial conditions $X_k(t_i) = 1$ and
$\dot{X_k}(t_i) = -i\omega(t_i)$.
The {\it spectral density} function $I(k,s,s')$ is defined as
\be
I(w,s,s') = \sum_k \d (\o -\o _k) \frac{c_k(s)c_k(s')}{2m_k(t_i) \o _k(t_i)}.
\te
One may decompose the influence kernel into its real and imaginary
parts, thus obtaining the dissipation and noise kernels:
\bea
\mu (s,s') &=& \frac{i}{2} \int_0^{\infty} dk I(k,s,s')
[X_k^{\ast}(s) X_k(s') - X_k^{\ast}(s') X_k(s)] \\
\nu (s,s') &=& \ha \int_0^{\infty} dk I(k,s,s')
[X_k^{\ast}(s) X_k(s') + X_k^{\ast}(s') X_k(s)].
\tea
By expressing the field as a collection of parametric oscillators,
it can be explicitly verified that the two approaches mentioned
above lead to the same result for the influence kernel $\zeta$. For
the purpose of calculating it in a specific case, we will find it
more convenient to use the second approach.

To study the thermal properties of the radiation measured by a
detector,  the influence kernel is compared to
that of a thermal bath of static oscillators each in a coherent state
\cite{HM2}:
\be \label{16-1}
\zeta = \int_0^\infty dk\;I_{\rm eff}(k,\Sigma)\;\left[C_k
(\Sigma)
\cos k\Delta-i\sin k\Delta\right]
\te
where
$$
\Sigma = (t + t')/2, ~~~ \Delta = t- t'
$$
and the function $C_k = \coth {\hbar k\over 2k_BT}$.
We will show in the specific cases discussed below that the
unknown function $C_k$ indeed has a $\coth$ form, and can then
deduce the temperature of the radiation seen by the detector. Here
$I_{\rm eff}(k,\Sigma)$ is the effective spectral density, also to  
be determined by formal manipulations of ($1.6$).  We can always
write $\zeta$ in this way since $\nu$ is even in
$\Delta$ while $\mu$ is odd.  By equating the real and imaginary
parts of the 
two forms of $\zeta$ and fourier inverting, we obtain
\bea
I_{\rm eff}C_k &=& {1\over\pi}
\int_{-\infty}^\infty d\Delta\;\cos k\Delta\;\nu(\Sigma,\Delta)
\label{18-1}\\
&&\nn \\
I_{\rm eff} &=& -{1\over \pi}
\int_{-\infty}^\infty d\Delta\;\sin k\Delta\;\mu(\Sigma,\Delta).
\label{16-5}
\tea

\noindent We will now consider various examples where $\zeta$ is
evaluated and shown to have, to zeroth order, a thermal form.
Higher-order corrections to $\zeta$ give a
{\it near-thermal} spectrum. In principle, the real and
imaginary parts of the influence kernel may be substituted
in the Langevin equation ($1.1$) to yield stochastic
near-thermal fluctuations of the detector coordinate $Q$.
This procedure will be demonstrated in the example of a
finite time uniformly accelerated detector (Sec. 3 below).
In this way, the methodology presented above describes
a {\it stochastic approach} to the problem of detector
response, as opposed to the usual perturbation theory
approach (where the perturbation parameter is $e^2$) involving the
calculation of detector transition probabilities. It should be
emphasized that equation ($1.1$) is exact for linear coupling
and does not involve a perturbation expansion in $e^2$ (for linear
systems, such an expansion is, in fact, unnecessary because they
are exactly solvable).

\newpage
\section{Asymptotically Uniformly-Accelerated Observer}
\label{non-uniformly-accelerated-observer}

We first consider the case of a non-uniformly accelerated monopole
detector in $1+1$ dimensions. For a general detector trajectory
$(x(\tau ), t(\tau )) $ parametrized by the proper time $\tau $,
it has been shown \cite{RHA} that the
 function $\zeta (\tau, \tau')$ is
\be
\zeta(\tau ,\tau^{\prime } ) \equiv  \nu +i\mu
    =  \frac{e^{2} }{2\pi } \int_{0}^{\infty} \frac{dk}{k}
e^{-ik(t(\tau )-t(\tau^{\prime } ))}
\cos k(x(\tau )-x(\tau^{\prime } )).
\te
Here
$e$ is the coupling constant of the detector to a massless scalar
field
(initially in its ground state). The initial state of the detector
is unspecified at the moment and would appear as a boundary
condition on the
equation of motion of the detector. Here,
however, we are primarily interested in the noise and dissipation
kernels themselves, as properties of the field, and not in the
state of the detector.

First, we note that the function $\zeta$ can be separated into
advanced and
retarded parts, in terms of the advanced and retarded null
coordinates
$v(\t )=t(\t )+x(\t )$ and $u(\t )=t(\t )-x(\t )$ respectively:
\bea
\z ^{a}(\t ,\t ') &=& \frac{e ^2}{4\pi}\int_{0}^{\infty}
\frac{dk}{k} e^{-ik(v(\t )-v(\t '))} \\ \nn
\z ^{r}(\t ,\t ') &=& \frac{e ^2}{4\pi}\int_{0}^{\infty}
\frac{dk}{k} e^{-ik(u(\t )-u(\t '))} \\ \nn
\z (\t ,\t ') &=& \z ^{a}(\t ,\t ') + \z ^{r}(\t ,\t ').
\tea

In the case when the detector is uniformly accelerated with
acceleration $a$, its trajectory is given by:
\be
v(\tau ) = \frac{1}{a} e^{a\tau } ;~~~
u(\tau ) = \frac{1}{a} e^{-a\tau }.
\te
Substitution of the above trajectory into equations (2.2) yields a
thermal, isotropic detector response at the Unruh temperature
$a/(2\pi)$ \cite{HM2,RHA}.
\subsection{Perturbation increasing with time}
The above analysis is now applied to the case of near-uniform
acceleration by
introducing a dimensionless $h$ parameter which measures the
departure from exact uniform acceleration:
\be
h =\frac{\dot{a} }{a^{2}} 
\te
where the overdot indicates derivative with respect to the proper
time. The trajectory of the detector is now chosen to be:
\be
v(\tau ) = \frac{1}{a(\t )} e^{\int a(\t )d\tau } ;~~~
u(\tau ) = \frac{1}{a(\t )} e^{-\int a(\t )d\tau }.
\te
One can expand $a(\t )$ in a Taylor series about the acceleration
at $\t =0$:
\be
a(\tau ) = a_{0} + \sum_{n=1}^{\infty } \frac{\tau ^{n} }{n! }
a^{(n)}_{0}
\te
where $a^{(n)}_{0} $ denotes the $n$-th derivative of $a$ at
$\tau =0$. We make
the assumption of ignoring second and higher derivatives of $a$.
This implies
\be
a(\tau ) = a_{0} + h_{0} \tau a_{0}^{2}
\te
where $h_{0} = \dot{a_0} /a_0^2 $.

Hereafter, we shall also make the further assumption of evaluating
quantities to
first order in $h_{0} $. In this approximation, $h =h_{0} $
to first order in $h_{0} $. Then there is no distinction between
$h $ and $h_{0} $ ($h $ is essentially
constant), and we can safely drop the subscript and work with $h $
alone. It should be noted that expanding quantities to first order
in $h $ actually involves expansion of quantities to first order
in $h \tau a_{0}$, and hence, for arbitrary trajectories, the final
results are
to be considered valid over time
scales $\tau $ such that $\tau \ll (h a_{0})^{-1} $. Alternatively,
equation
($2.7$) can be taken to define a family of trajectories for which
this analysis applies.

Using the linearized form of $a(\tau )$, one can now obtain the
trajectory explicitly, to first order in $h $. The result is:
\bea
v(\tau ) &=& a_{0}^{-1} e^{a_{0}\tau}\left(1 +
h\t a_0(\frac{a_0\t }{2}-1)\right) \nn \\
u(\tau ) &=& -a_{0}^{-1} e^{-a_{0}\tau}\left(1 -
h\t a_0(\frac{a_0\t }{2}+1)\right)
\tea

One also finds, to first order in $h $,
\bea
e^{-ik(v(\tau )-v(\tau^{\prime } ))}
& = & e^{-\frac{2ik}{a_{0}} e^{a_{0}\Sigma } \sinh
(\frac{a_{0}\Delta }{2} )}
\left[1-ikh e^{a_{0}\Sigma }
\left((\frac{a_0\D ^2}{4}+a_0\Sigma^2-2\Sigma)
\sinh (\frac{a_{0}\Delta }{2})\right. \right.  \nn \\
&   &+ \left. \left.\Delta(a_0\Sigma-1) \cosh
(\frac{a_{0}\Delta }{2} )\right)\right]  \\
e^{-ik(u(\tau )-u(\tau^{\prime } ))}
& = & e^{-\frac{2ik}{a_{0}} e^{-a_{0}\Sigma } \sinh
(\frac{a_{0}\Delta }{2} )}
\left[1+ikh e^{-a_{0}\Sigma }
\left((\frac{a_0\D ^2}{4}+a_0\Sigma^2+2\Sigma)
\sinh (\frac{a_{0}\Delta }{2})\right. \right.  \nn  \\
&   &-\left. \left. \Delta(a_0\Sigma+1) \cosh
(\frac{a_{0}\Delta }{2} )\right)\right]
\tea
where $\Delta =\tau -\tau^{\prime } $, $\Sigma =\frac{1}{2}
(\tau +\tau')$. 

Using the identities \cite{Ang}
\be \label{2-10}
e^{-\frac{2ik}{a_{0}} e^{-a_{0}\Sigma }\sinh (\frac{a_{0}\D}{2} )}
= \frac{4}{\pi } \int_{0}^{\infty } d\nu K_{2i\nu }(\frac{2k}{a_{0}}
e^{-a_{0}\Sigma })[\cosh (\pi \nu )\cos (\nu a_{0}\Delta ) -
i\sinh (\pi \nu )\sin (\nu a_{0}\Delta )]
\te
and
\be \label{2-11}
\mid \Gamma(i\nu)\mid^2 = \frac{\pi}{\nu \sinh \pi \nu} ;~~~
\mid \Gamma(\frac{1}{2}+i\nu)\mid^2 = \frac{\pi}{\cosh \pi \nu}
\te
one finally obtains, after some simplification,
\bea
\zeta^{a}(\tau ,\tau^{\prime } ) = \frac{e^{2}}{4\pi }
\int_{0}^{\infty } \frac{dk}{k} [\coth (\frac{\pi k }{a_{0}} )
\cos (k \Delta )(1+ h \Gamma_{1}) - i\sin (k \Delta ) ] \\
\zeta^{r}(\tau ,\tau^{\prime } ) = \frac{e^{2}}{4\pi }
\int_{0}^{\infty } \frac{dk}{k} [\coth (\frac{\pi k }{a_{0}} )
\cos (k \Delta )(1+ h \Gamma_{2}) - i\sin (k \Delta ) ]
\tea
with
\bea
\Gamma_{1} &=& -k \tan (k \Delta )\tanh^{2} (\frac{\pi k }{a_{0}})
\left[(\frac{a_0\D ^2}{4}+a_0\Sigma^2-2\Sigma)
\sinh (\frac{a_{0}\Delta }{2}) + \Delta(a_0\Sigma-1) \cosh
(\frac{a_{0}\Delta }{2})\right] \nn \\
\Gamma_{2} &=& k \tan (k \Delta )\tanh^{2} (\frac{\pi k }{a_{0}})
\left[(\frac{a_0\D ^2}{4}+a_0\Sigma^2+2\Sigma)
\sinh (\frac{a_{0}\Delta }{2}) \right.\nn \\
& &- \left. \Delta(a_0\Sigma+1) \cosh
(\frac{a_{0}\Delta }{2})\right]
\tea
The advanced and retarded parts of $Re(\zeta)$ being unequal, the
noise
is anisotropic. Adding expressions ($2.12$) and ($2.13$), we have
\be
\zeta(\tau ,\tau^{\prime } ) = \frac{e^{2}}{2\pi } \int_{0}^{\infty }
\frac{dk}{k} [\coth (\frac{\pi k }{a_{0}} )
\cos (k \Delta )(1+ h \Gamma) - i\sin (k \Delta ) ]
\te
where
\be
\Gamma = \frac{\Gamma_1 + \Gamma_2}{2} = k\Sigma\tan (k \Delta )
\tanh^{2}(\frac{\pi k }{a_{0}})(2\sinh \frac{a_{0}\D}{2} -
a_0\Delta \cosh\frac{a_{0}\D}{2}).
\te
The noise experienced by the detector is thus identical to the noise
experienced in a heat bath, with a small correction, $\Gamma$.
The accelerated detector therefore has a near-thermal response at the
Unruh temperature $a_0/(2\pi)$ with an order $h$ correction which
increases with time.

\subsection{Perturbation exponentially decreasing with time}

\noindent We will now consider a trajectory for the accelerated
detector which
exponentially approaches the uniformly accelerated trajectory at
late times.
This trajectory, in null coordinates, has the form
\be
v(\t ) = a_{0}^{-1}e^{a_{0}\t }(1+\a e^{-\g \t });~~~u(\t ) =
-a_{0}^{-1}e^{-a_{0}\t }(1+\a e^{-\g \t }).
\te
In this case, the magnitude of the proper acceleration is, to first
order in $\a $,
\be
a(\t ) =  a_{0}\{1+\a e^{-\g \t }(1+\frac{\g ^2}{a_0^2})\} +
O(\a ^2).
\te
The influence kernel is obtained in a manner similar to the
treatment of the
previous subsection. Here, we get
\bea
\zeta^{a}(\tau ,\tau^{\prime } ) = \frac{e^{2}}{4\pi }
\int_{0}^{\infty } \frac{dk}{k} [\coth (\frac{\pi k }{a_{0}} )
\cos (k \Delta )(1+ \a \Gamma_{1}) - i\sin (k \Delta ) ] \\
\zeta^{r}(\tau ,\tau^{\prime } ) = \frac{e^{2}}{4\pi }
\int_{0}^{\infty } \frac{dk}{k} [\coth (\frac{\pi k }{a_{0}} )
\cos (k \Delta )(1+ \a \Gamma_{2}) - i\sin (k \Delta ) ]
\tea
with
\bea
\Gamma_{1} &=& -2ka_0^{-1}e^{-\g \Sigma }\sinh
\frac{(a_0-\g )\D }{2}
\tan (k \Delta )\tanh^{2} (\frac{\pi k }{a_{0}}) \nn \\
\Gamma_{2} &=& -2ka_0^{-1}e^{-\g \Sigma }\sinh
\frac{(a_0+\g )\D }{2}
\tan (k \Delta )\tanh^{2} (\frac{\pi k }{a_{0}})
\tea
The noise
is again seen to be anisotropic. Adding $\zeta^{a}$ and $\zeta^{r}$,
we have
\be
\zeta(\tau ,\tau^{\prime } ) = \frac{e^{2}}{2\pi } \int_{0}^{\infty }
\frac{dk}{k} [\coth (\frac{\pi k }{a_{0}} )
\cos (k \Delta )(1+ \a \Gamma) - i\sin (k \Delta ) ]
\te
where
\be
\Gamma = \frac{\Gamma_1 + \Gamma_2}{2} = -2ka_0^{-1}e^{-\g \Sigma }
\sinh \frac{a_0\D }{2}
\cosh \frac{\g \D }{2}\tan (k \Delta )\tanh^{2}
(\frac{\pi k }{a_{0}}).
\te
In this case, the correction to the thermal spectrum is
exponentially suppressed
at late times. This feature will distinguish the behavior of
quantum fields
in the vicinity of a moving mirror and a collapsing mass,
as shown in later sections.

\newpage
\section{Finite-Time Uniformly-Accelerated Detector}

In this section, we consider a detector trajectory which is a
uniformly
accelerated one for a finite interval of time $(-t_0, t_0)$.
Before and after this interval,
the trajectory is taken to be inertial, at uniform velocity. To
ensure
continuity of the proper time along this trajectory, the velocity
of the
detector is assumed to vary continuously at the junctions $ \pm t_0$.

With these constraints, the trajectory is chosen to be
\bea
x(t) &=& x_{0}^{-1} (a^{-2}-t_{0}t),  ~~~~  t<-t_0 \nn  \\
&=& (a^{-2} + t^{2})^{\ha },  ~~~~  t>-t_0, ~t<t_0   \nn \\
&=& x_{0}^{-1} (a^{-2}+t_{0}t),  ~~~~  t>t_0.
\tea
The trajectory is symmetric under the interchange $t\rightarrow -t$.
$a$ is the magnitude of the proper acceleration during the uniformly
accelerated interval $(-t_0, t_0)$ of Minkowski time and $x_0$ is
the position of the detector at
time $t_0$. $x_0$ and $t_0$ are related by $x_0^2 - t_0^2 = a^{-2}$.
Before the uniformly accelerated interval, the detector has a
uniform velocity
$-t_0/x_0$ (we have chosen units such that $c=1$; if one
keeps factors of $c$, the velocity is $-c^2t_0/x_0$)
and after this interval, its velocity is $t_0/x_0$.
This trajectory
thus describes an observer traveling at constant velocity, then
turning around
and traveling with the same speed in the opposite direction.
The ``turn-around'' interval corresponds to the interval of uniform
acceleration.

We may also define null coordinates $u = t - x$ and $v = t + x$. In
terms of
these, the time at which the trajectory crosses the future horizon
$u=0$ of the uniformly accelerated interval is
$t_{H} = -(a^2u_0)^{-1}$.

If we choose to parametrize the trajectory by the proper time $\t $,
it can be
expressed as (with the zero of proper time chosen at $t=0$)
\bea
u(\t ) &=& \q (-\t _0-\t )v_0\{a(\t + \t _0) - 1\} -
a^{-1}\q (\t _0 + \t )
\q (\t _0 - \t ) e^{-a\t }  \nn \\
& &+ \q (\t -\t _0)u_0\{ 1+a(\t _0 - \t )\} \\
v(\t ) &=& -\q (-\t _0-\t )u_0\{a(\t + \t _0) + 1\} +
a^{-1}\q (\t _0 + \t )
\q (\t _0 - \t ) e^{a\t }  \nn \\
& &+ \q (\t -\t _0)v_0\{ 1-a(\t _0 - \t ) \}
\tea
where $\pm\t _0$ is the proper time of the trajectory when it exits (enters)
the uniformly accelerated phase. It satisfies the relations
\bea
v_0 &\equiv& \tz + \xz = a^{-1}e^{a\t _0} \nn  \\
u_0 &\equiv& \tz - \xz = -a^{-1}e^{-a\t _0}.
\tea
Another convenient definition is the horizon - crossing proper time
$\pm \tah =\pm (a^{-1}+\taz )$.

The function $\zeta(\t ,\t ')$ can be
found in a standard way. If both points lie on the inertial
sector of
the trajectory, it has the usual zero-temperature form in
two-dimensional
Minkowski space. If both points lie on the uniformly
accelerated sector, it
has a finite temperature form exhibiting the Unruh temperature.
It is therefore straightforward to obtain the following:

\noindent If $\underline{\t , \t '>\t _0}$ or
$\underline{\t , \t '<-\t _0}$,
\be
\zeta(\t ,\t ') = \frac{e^2}{2\pi}\intf \frac{dk}{k}e^{ik(\t ' -\t )}.
\te

\noindent If $\underline{-\t _0<\t , \t '<\t _0}$,
\be
\zeta(\t ,\t ') = \frac{e^2}{2\pi}\intf \frac{dk}{k}
\{\coth(\frac{\pi k}{a}) \cos k(\t ' -\t ) + i\sin k(\t ' -\t )\}.
\te

\noindent Also, if $\underline{\t <-\t _0, \t '>\t _0}$,
\be
\zeta(\t ,\t ') = \frac{e^2}{2\pi}\intf \frac{dk}{k}
\cos k((\t '+\t )\tanh(a\t _0))
e^{ik(\t ' -\t + 2(a^{-1}\tanh(a\t _0)-\t _0))}.
\te
Of interest
is this function evaluated for one point on the inertial sector
and the other
on the uniformly accelerated sector. We will show that this
function has a thermal form if the point on the inertial sector
is sufficiently close to $(t_0, x_0)$ and departs smoothly from
the thermal form away from it. It is
also found that the horizons of the uniformly accelerated sector
(which are {\it not} horizons for the entire trajectory) are the
points where the near-thermal expansion breaks down.

Consider, for example, the case when $\underline{-\taz < \t ' < \taz}$
and $\underline{\t < -\taz}$.
Then the function $\zeta$ is expressed as
\be
\zeta(\t ,\t ') = \frac{e^2}{4\pi}\intf \frac{dk}{k}
\{e^{-ik(a^{-1}e^{-a\t '}
+ \vz (a(\t +\taz ) -1))} + e^{ik(a^{-1}e^{a\t '} + \uz
(a(\t +\taz ) +1))}\}.
\te
Introducing the Fourier transforms
\bea
e^{\frac{ik}{a}e^{a\t }} &=& \frac{1}{2\p a} \intin d\o
e^{i\o \t } \G
(-\frac{i\o }{a}) (\frac{k}{a})^{\frac{i\o }{a}}
e^{\frac{\p \o }{2a}} ,~~~ k>0
\nn \\
e^{-\frac{ik}{a}e^{-a\t }} &=& \frac{1}{2\p a} \intin d\o
e^{i\o \t } \G
(\frac{i\o }{a}) (\frac{k}{a})^{-\frac{i\o }{a}}
e^{\frac{\p \o }{2a}} ,~~~ k>0
\tea
we obtain, after some simplification,
\bea
\zeta(\t ,\t ') &=& \frac{e^2}{4\pi}\intf \frac{dk}{k}
\left\{\cos k\left(\t ' +\taz +
\frac{1}{a}\ln (1-a(\t +\taz))\right)\coth (\frac{\pi k}{a})
\right.\nn \\
&+& i\sin k\left(\t ' +\taz +\frac{1}{a}\ln (1-a(\t +\taz))
\right) \nn\\
&+& \cos k(\t ' +\taz -\frac{1}{a}\ln \mid a(\t +\tah)\mid)
\left(\coth (\frac{\pi k}{a}) \q (\tah +\t ) + \q (-\tah -\t )
\right) \nn\\
&+& \left. i\sin k(\t ' +\taz - \frac{1}{a} \ln \mid a(\t +\tah)\mid)
\q (\tah +\t )\right\}.
\tea
If we further restrict our attention to the case
$\underline{\t > -\tah }$, 
i.e. both
points lie inside the Rindler wedge, the above expression simplifies
to  the following :
\bea
\zeta(\t ,\t ') &=& \frac{e^2}{2\pi}\intf \frac{dk}{k}
\cos\left(\frac{k}{2a}\ln (1-a^2 (\t +\taz )^2)\right)\times \nn\\
& &\left\{\coth(\frac{\pi k}{a})
 \cos k\left(\t '+\taz + \frac{1}{2a}
\ln (1-2\frac{\t +\taz }{\t +\tah })\right) \right. \nn \\
&+& \left. i\sin k\left(\t '+\taz + \frac{1}{2a}
\ln (1-2\frac{\t +\taz }{\t +\tah })\right)\right\}.
\tea 
It is clear from this expression that an exact thermal spectrum is
recovered in the limit of $\t \rightarrow -\taz$, as expected.
Suppose we now define $\t +\taz \equiv\e $ as the time difference
between the proper time
$\t $ and the proper time of entry into the accelerated phase,
$-\taz $. Then $a\e $ will be the appropriate dimensionless
parameter characterizing a near-thermal expansion. Note that $\e <0$.

{}From the above expression for $\zeta$, we find that there is no
correction
to the thermal form of $\zeta(\t ,\t ')$ to first order in $\e $.
This can be understood from the fact that the coordinate difference
between the point $\t =-\taz -\e $ and a corresponding point on a
globally
uniformly accelerated trajectory with the {\it same} proper time
is of order $\e ^2$. Indeed, we may define Rindler coordinates
$(\psi, \eta)$ on the right Rindler wedge by
$v=\psi^{-1}e^{\psi \eta}$ and $u=-\psi^{-1}e^{-\psi \eta}$.
Then the Rindler coordinates for the point $\t =-\taz -\e $
on the trajectory we consider are found to be $\psi = a +
{\cal O}(\e ^2)$
and $\eta =-\taz -\e + {\cal O}(\e ^2)$, which are exactly the
coordinates,
to order $\e $, of a corresponding point with the same proper
time on a
globally uniformly accelerated trajectory with acceleration $a$.
It is thus
no surprise that the spectrum is exactly thermal up to order $\e $.

Furthermore, it can be shown in a straightforward way from the above
expression that the spectrum is also thermal up to ${\cal O}(\e ^2)$,
although the above-mentioned coordinate difference does have terms of
order $\e ^2$. Then the first correction to the thermal spectrum
is of order $\e ^3$ and has the form
\be
\zeta(\t ,\t ') = \frac{e^2}{2\pi}\intf \frac{dk}{k}
\{\coth(\frac{\pi k}{a})
\cos k(\t '-\t + \frac{a^2\e ^3}{3})
 - i\sin k(\t '-\t + \frac{a^2\e ^3}{3})\} + {\cal O}(\e ^4).
\te

The validity of such a near-thermal expansion is
characterized by the requirement that $\mid a\e \mid$ is small. This
translates to $-1<a(\t +\taz )$ or equivalently, $\t >-\tah$.
The expansion thus breaks down for $\t < -\tah$, for which case the
two-point function
may be called strictly non-thermal. This is the case when one of
the points
lies outside the right Rindler wedge while the other point is
still inside
it. The two-point function in such a situation will contain
non-trivial correlations across the Rindler horizon, as was
pointed out before~\cite{RHA}.

The response of the detector is governed by the Langevin equation
($1.1$). This
equation may be formally integrated to yield
\be
\langle Q(\t )Q(\t ')\rangle = \frac{\hbar}{\Omega^2}
\int_{-\infty}^{\t }ds
\int_{-\infty}^{\t '}ds'\nu(s, s') e^{-\gamma(\t -s)}
e^{-\gamma(\t '-s')}
\sin\Omega(\t -s)\sin\Omega(\t '-s')
\te
where $\Omega = (\Omega_0^2 - \gamma^2)^{\frac{1}{2}}$,
$\Omega_0$ is the
natural frequency of the internal detector coordinate and
$\gamma = e^2/4$
is the dissipation constant arising out of the detector's
coupling to the
field. The double integral in the above equation may be computed by
splitting each integral into a part which lies completely in the
uniformly
accelerated sector and parts which lie in the inertial sectors. For
example,
suppose we wish to compute the above correlation function for the
case $-\taz < \t ,\t '<\taz$, i.e. both points lie in the uniformly
accelerated sector. Then each integral can be split into two parts
($\int_{-\infty}^{\t } = \int_{-\infty}^{-\taz} +
\int_{-\taz}^{\t }$)
and the resulting
double integral therefore has four terms:
\be
\langle Q(\t )Q(\t ')\rangle = F_1 + F_2 + F_3 + F_4. 
\te
Writing $\nu \equiv Re(\zeta)$, we obtain, after straightforward 
manipulations,
\bea
F_1 &\equiv& \frac{2\hbar \gamma}{\pi \Omega^2} Re
\int_{-\infty}^{-\taz }ds
\int_{-\infty}^{-\taz}ds' \intf \frac{dk}{k} e^{ik(s'-s)}
e^{-\gamma(\t -s)}
e^{-\gamma(\t '-s')}\times\nn  \\
& &\sin \Omega(\t -s) \sin \Omega(\t '-s') \nn \\
&=& \frac{\hbar \gamma}{\pi \Omega^2} e^{-\gamma(\t +\t '+2\taz )}
\intf \frac{dk}{k}[(\gamma^2 -k^2 +\Omega^2)^2 + 4\gamma^2k^2]^{-1}
\times \nn \\
& &\{(\Omega^2 +\gamma^2 + k^2 )\cos \Omega(\t -\t ') + (\Omega^2 -
\gamma^2 - k^2)\cos \Omega (\t +\t '+2\taz ) \nn\\
& & + 2\gamma\Omega\sin
\Omega (\t +\t '+2\taz )\},
\tea
and
\bea
F_4 &\equiv& \frac{2\hbar \gamma}{\pi \Omega^2} Re
\int_{-\taz}^{\t }ds
\int_{-\taz}^{\t '}ds' \intf \frac{dk}{k} e^{ik(s'-s)}\coth
(\frac{\pi k}{a})
e^{-\gamma(\t -s)}e^{-\gamma(\t '-s')}\times\nn\\
& &\sin \Omega(\t -s) \sin \Omega(\t '-s')
\nn \\
&=& \frac{\hbar \gamma}{\pi \Omega^2}
e^{-\gamma(\t +\t '+2\taz )} \intf
\frac{dk}{k}\coth (\frac{\pi k}{a})
[(\Omega^2 +\gamma^2 -k^2 )^2 + 4\gamma^2k^2]^{-1}
\times \nn \\
& &\{(\gamma^2 + k^2 +\Omega^2)\cos \Omega(\t -\t ')
+ (\Omega^2 -\gamma^2 - k^2)\cos \Omega (\t +\t '+2\taz )
\nn \\
& &+2\gamma\Omega(\sin \Omega (\t +\t '+2\taz )-\sin \Omega (\t
+\taz ) - \sin \Omega (\t '+\taz )) \nn \\
& &-\Omega^2 (\cos \Omega(\t +\taz ) + \cos\Omega(\t '+\taz)
-1)\},
\tea
where $Re$ stands for the real part.

The functions $F_2$ and $F_3$, in which one of the integration
variables runs over the inertial sector and the other over
the uniformly accelerated sector, are difficult to evaluate.
We shall simply express them here in the following form :
\bea
F_2 &=& \frac{\hbar \gamma}{\pi \Omega^2}e^{-\gamma(\t +\t ')}
 Re \int_{-\infty}^{-\taz }ds e^{\gamma s}\sin \Omega(\t -s)
 \times \nn \\
& & \intf \frac{dk}{k}
[e^{ik(a\uz s +\uz (1+a\taz ))}A_1(k;\t ')
+ e^{-ik(a\vz s -\vz (1-a\taz ))}A_2(k;\t ')]\\
F_3 &=& \frac{\hbar \gamma}{\pi \Omega^2}e^{-\gamma(\t +\t ')}
 Re \int_{-\infty}^{-\taz }ds e^{\gamma s}\sin \Omega(\t' -s)
 \times \nn \\
& &\intf \frac{dk}{k}
[e^{-ik(a\uz s +\uz (1+a\taz ))}A_1(k;\t )
+ e^{ik(a\vz s -\vz (1-a\taz ))}A_2(k;\t )]
\tea
where the functions $A_1$ and $A_2$ are
\bea
A_1(k;s) &=& \int_{-\taz}^{s}ds'e^{ik a^{-1}e^{as'}}
e^{\gamma s'}\sin \Omega(s-s')   \nn\\
A_2(k;s) &=& \int_{-\taz}^{s}ds'e^{-ik a^{-1}e^{-as'}}
e^{\gamma s'}\sin \Omega(s-s').
\tea
Similarly, if one wishes to compute the detector correlation
function for two points in the late inertial sector ($\t ,\t
' > \taz $), then one has nine terms similar in form to the
ones displayed above.

\newpage
\section{Moving mirror in Minkowski space}
\label{moving-mirror-in-minkowski-space}

In this section, we treat the motion of a mirror following a
trajectory $z(t)$ in
Minkowski space. A massless scalar field $\phi$ is coupled to the
mirror via a reflection boundary condition. It obeys the
Klein-Gordon equation
\be
{\partial^2 \phi\over \partial t^2} - {\partial^2 \phi\over
\partial x^2} = 0
\te
subject to the boundary condition
\be \label{fullingdavies2.3}
\phi(t,z(t)) = 0.
\te
For a general mirror path this equation is difficult to solve;
however we can exploit the invariance of the wave equation
under a conformal transformation
to change to simpler coordinates.  We follow the treatment 
of~\cite{FulDav}.  To this end, we introduce
a transformation between the null coordinates $ u, v$ and
$\overline u, \overline v$ defined as
\bea \label{fullingdavies2.4}
u &=& t-x, ~~ v= t+x , \nn \\
u &=& f(\overline u), ~~v = \overline v
\tea
The function $f$ is chosen such that the mirror trajectory is
mapped to
$\overline z = 0$.  To do this, we relate the two sets of
coordinates as follows:
\bea
t &=& {1\over 2}[\;\overline v + f(\overline u)] \nn\\
x &=& {1\over 2}[\;\overline v - f(\overline u)]
\tea
On the mirror path, setting $\overline z = 0$ means that the
trajectory can be expressed as
\be \label{201-4}
{1\over 2}[\;\tbar-f(\tbar)] = z\left({1\over 2}
[\;\tbar+f(\tbar)]\right)
\te
which allows $f$ to be implicitly determined.  In the new
coordinates the wave 
equation is unchanged, however it now has a time independent
boundary condition, meaning the mirror is static, while the
detector moves along
some more complicated path.  Thus the wave equation with boundary
condition can easily be solved to give
\be \label{200.1-0}
\phi(\overline t,\overline x) = \int_0^\infty (2\pi k)^{-1/2}
\sin k\overline x \;e^{-ik\overline t}\;dk
\te
where the mode functions are orthonormal in the Klein-Gordon
inner product.
In  these barred coordinates, $\zeta$ is proportional to the two
point function in the presence of a {\it static} reflecting boundary
at $\overline z=0$.  

Also, in these coordinates, the time dependent modes of the field
are just
exponentials.  That is, they can be described by oscillators with
unit mass
and frequency $k$.  So $X_k(\overline t)$ is a solution to the
oscillator equation (1.4), and by satisfying the initial conditions
$X_k(0)=1, X_k'(0) = -ik$ we
obtain
\be \label{200.1-1}
X_k(\overline t) = e^{-ik\overline t}
\te

We now consider a detector placed in the vicinity of the mirror. The
spectral density function $I$ is determined by the path of the
detector and its coupling to the field. Denoting the
detector position by $r(t)$ and the field modes by $q_k(t)$ and
assuming the monopole interaction
\bea
L_{\rm int} &=& -\int eQ\phi(\overline t,\overline x)\;
\delta(\overline r- \overline x)\;d\overline x \nn \\
&=& -eQ\phi(\overline t,\overline r) \nn \\
&=& -\int eQq_k(\overline t)\;\sin k\overline r \;dk,
\tea
we have
\bea 
I(k,\overline t,\overline t') &=& \int {dk_n\over 2k_n}
\delta(k-k_n) e^2 \sin k_n \overline r(\overline t)
\sin k_n \overline r(\overline t') \nn \\
&=& {e^2\over 2k}\sin k \overline r(\overline t)
\sin k \overline r(\overline t'). \label{200.1-5}
\tea
Defining $\overline u = \overline t - \overline r(t)$ and
$\overline v = \overline t + \overline r$, we can now express
the function $\zeta$ as
\be \label{212-2}
\zeta =-{e^2\over 8\pi} \int_0^\infty {dk\over k}\;
\left[e^{ik( \overline u' -
\overline u)}-e^{ik(\overline u'-\overline v)}-
e^{ik(\overline v'-\overline u)}+ e^{ik(\overline v' -
\overline v)}\right]
\te
Since only the outgoing modes have reflected off the mirror,
only the outgoing part of the correlations $\zeta$ will give
appropriate
thermal behavior. Thus, from now on, we focus on the correlation
\be
\zeta_{uu} = -{e^2\over 8\pi} \int_0^\infty {dk\over k}\;
e^{ik( \overline u' - \overline u)}.
\te
It remains to evaluate the above function. To do this, we specify
the function
$f$ by considering a specific mirror trajectory.
A convenient choice of the mirror path is the following:
\be \label{birrell-davies4.51}
z(t) = -t - Ae^{-2\kappa t} + B
\te
for $A$, $B$, $\kappa$ positive.
This path possesses a future horizon in the sense
that there is a last ingoing ray which the mirror will reflect;
all later rays never catch up with the mirror and so are not
reflected.  It is this aspect
which enables the moving mirror to emulate a black hole.  
Eq.\ \eqn{201-4} can now be solved to give
\be
f(\overline t) = -\overline t -{1\over\kappa} \ln
{B-\overline t\over A}.
\te
In the late time limit ($\overline t \simeq B$), we consider the
following ansatz for $f^{-1}$
\be \label{201-6}
f^{-1}(x) \simeq B-Ae^{-\kappa (B+x)} + \alpha
\te
where $\alpha$ is taken to be small in the sense that terms of
order $\alpha^2$ are ignored. In this approximation, one finds
\be
\alpha = -\kappa A^2 e^{-2\kappa(B+x)}
\te
and the transformation from barred to unbarred coordinates becomes
\be
\overline u = B - Ae^{-\kappa (B+u)} - \kappa A^2 e^{-2\kappa(B+u)}
\te
plus terms of higher powers in $e^{-\kappa (B+u)}$.

We now need an explicit form for the detector trajectory $u(t)$
since this is what appears in the function $\zeta $.
 Choosing it to be inertial,
 we have $r(t) = r_\ast + wt$, which gives $u(t) = t(1-w)-r_\ast$.
In terms
of the proper time of the detector, this becomes $u(\tau) = \tau
\sqrt{(\frac{1-w}{1+w})} - r_\ast$.

Defining the sum and difference $\Sigma = \ha (\tau + \tau')$ and
$\Delta = \tau -\tau'$, and $z = \sqrt{(\frac{1-w}{1+w})}$,
we obtain
\be
\overline u' - \overline u = -2Ae^{-\kappa(B+r_\ast +\Sigma\nu)}
\sinh (\frac{\kappa \nu \Delta}{2}) - 2\kappa A^2
e^{-2\kappa(B+r_\ast +\Sigma\nu)}\sinh (\kappa \nu \Delta).
\te
This is substituted in $\zeta_{uu}$, and, after some simplification
we obtain the near-thermal form
\be
\zeta_{uu}(\tau ,\tau^{\prime } ) = \frac{e^{2}}{8\pi }
\int_{0}^{\infty } \frac{dk}{k} [\coth (\frac{\pi k }{\kappa\nu} )
\cos (k \Delta )(1+ h \Gamma) - i\sin (k \Delta ) ]
\te
with
\be
\Gamma = -2k e^{-\kappa(B+r_\ast +\Sigma \nu)} \tanh^2
(\frac{\pi k}{\kappa\nu}) \tan k\Delta.
\te
Thus a thermal detector response, at the temperature
$\frac{\kappa}{2\pi}$,
Doppler-shifted by a factor $\nu$ depending on the speed of the
detector, is observed, with a correction that exponentially decays
to zero at late times.

\newpage
\section{Collapsing Mass in Two Dimensions}

In this section we study radiance from a collapsing mass, analogous
to
the moving mirror model. We essentially follow the method of \cite
{BirDav}, but using stochastic analysis, and generalizing it to
include
higher order terms in the various Taylor expansions involved, thus
exhibiting the near-thermal properties of detector response.

We will exploit the conformal flatness of two dimensional spacetime
in the subsequent analysis. Outside the body the metric is expressed
as
\be
ds_o^2 = C(r)du\,dv
\te
where $u$, $v$ are the null coordinates
\bea
u &=& t-r^{\ast} + R_0^{\ast}  \\ \nn
v &=& t+r^{\ast} + R_0^{\ast}
\tea
and $r^\ast$ is the Regge-Wheeler coordinate:
\be \label{4-3}
r^\ast = \int^{r} \frac{dr'}{C(r')}
\te
with $R_0^{\ast}$ being a constant. The metric outside the body is
thus assumed to be static in order to mimic the four dimensional
spherically
symmetric case (for which Birkhoff's theorem holds). The point at
which the
conformal factor $C=0$ represents the horizon, and the asymptotic
flatness
condition is imposed by $C\rightarrow 1$ as $r\rightarrow \infty$.

On the other hand, the metric inside the ball is for now assumed
to be a completely general conformally flat metric:
\be
ds_i^2 = A(U,V)dUdV
\te
with
\bea
U &=& \tau -r + R_0  \nn \\
V &=& \tau -r - R_0
\tea
and $R_0$ and $R_0^\ast$ are related in the same way as $r$ and
$r^\ast$.
The surface of the collapsing ball will be taken to follow the
worldline
$r=R(\tau)$, such that, for $\tau <0$, $R(\tau)=R_0$. Thus, at the
onset of collapse, $\tau = t =0$, $U=V=u=v=0$ on the surface of
the ball.

We will let the two sets of coordinates be related by the
transformation equations
\bea
U &=& \alpha(u)  \nn \\
v &=& \beta(V).
\tea
The functions $\alpha$ and $\b $ are not independent of each other
because one
coordinate transformation has already been specified by equation
\eqn{4-3}.

Without as yet determining the precise form of $\a $ and $\b $,
we will consider a massless scalar field $\phi$ propagating in this
spacetime subject
to a reflection condition $\phi(r=0,\tau)=0$. Such a field
propagates in a similar fashion to the field in the vicinity of a
moving mirror. To make this
explicit, we introduce a new set of barred coordinates
\bea
\overline u &=& \b [\a (u) - 2R_0]  \nn \\
\overline v &=& v.
\tea
In terms of these, we also define the coordinates
$\overline r=\ha(\overline v
- \overline u)$, and $\overline t=\ha(\overline v + \overline u)$.

These new coordinates have the properties : $a)\,  r=0
\Rightarrow \overline r = 0$,
and $b)$ the field equations have incoming mode solutions of the
form $e^{ikv}$.
Thus the left-moving parts of the correlation functions of the `in'
vacuum
defined in terms of barred coordinates are identical to those of
the vacuum
defined with respect to unbarred coordinates.

Keeping these properties in mind, we may expand the field in terms
of standard
modes obeying the reflection boundary condition (by conformal
invariance of the massless scalar field equation) as
\be
\phi(\overline r, \overline t) = \sqrt{\frac{2}{L}} \sum_{k>0}
\overline q_k (\overline t) \sin k\overline r,
\te
just as in the moving mirror case.

We now consider a detector placed outside the collapsing ball at
fixed $r$ (or
$r^\ast$), namely $r=r_0$ (or $r^\ast = r_0^\ast$). The interaction
between
detector and field is described by the interaction Lagrangian
\be
L_{int} = -\epsilon Q\phi(\overline s, \overline r),
\te
where
\bea
\overline r &=& \ha (v-\b (\a (u) - 2R_0)) \\ \nn
&=& \ha (t+r_0^{\ast}-R_0^{\ast}-\b (\a
(t-r_0^{\ast}+R_0^{\ast})-2R_0)) \\ \nn
\overline s &=& \ha (t+r_0^{\ast}-R_0^{\ast}+\b
(\a (t-r_0^{\ast}+R_0^{\ast}) -2R_0))
\tea
and $Q$ is the internal detector coordinate.

The influence kernel $\zeta$,
 due to a reflection condition at $\overline r=0$, has the same
form as the moving
mirror case, in barred coordinates. Its outgoing part
is therefore given by
\be
\zeta_{uu} = \frac{e^2}{8\pi}\int_0^{\infty} \frac{dk}{k}
e^{ik(\overline u' - \overline u)}
\te
where
\be
\overline u = \overline s - \overline r = \b (\a
(t-r_0^{\ast}-R_0^{\ast}) -2R_0)
\te
and $\overline u'$ is the same function of $t'$.

We will now determine the functions $\a $ and $\b $ and show that,
to zeroth order in an appropriate parameter, $\overline u$ is an
exponential function
of $t$, and thus $\zeta_{uu}$ has a thermal form. The correction to
the exponential form, obtained by including higher order terms, will
lead to a near-thermal spectrum.

To determine $\a $ and $\b $ we match the interior and exterior
metrics at
the collapsing surface $r=R(\t )$. Then we have
\bea
\a '(u) \equiv \frac{dU}{du} &=& -C\frac{(1-\dot{R})}{\dot{R}}
[1+(1+\frac{AC}{\dot{R}^2}(1-\dot{R}^2))^{\ha }]^{-1} \\
\b '(V) \equiv \frac{dv}{dV} &=& C^{-1}\frac{\dot{R}}{1+\dot{R}}
[1-(1+\frac{AC}{\dot{R}^2}(1-\dot{R}^2))^{\ha }]
\tea
where $\dot{R}=\frac{dR}{d\t }$.

Now we expand these quantities about the horizon. We recall the
definition of
the horizon radius $R_h$ as $C(R_h)=0$. We may further define
$\t _h$ as
$R(\t _h)=R_h$. Then we obtain the following Taylor expansions:
\be
R(\t ) = R_h + \nu (\t _h -\t ) + \b (\t _h - \t )^2 + \cdots
\te
where $\nu = -\dot{R}(\t _h)$, $\b = \ha \ddot{R}(\t _h)$, and
\bea
C &=& \frac{\partial C}{\partial r}\mid_{R_h}(R-R_h) + \ha
\frac{\partial^2 C}{\partial r^2}\mid_{R_h}(R-R_h)^2 + \cdots   \\ \nn
&=& 2\kappa \nu(\t _h-\t ) + (2\kappa \beta + \gamma \nu^2)
(\t _h - \t )^2 + \cdots
\tea
where $\kappa = \ha \frac{\partial C}{\partial r}\mid_{R_h}$, the
surface gravity, and
$\gamma = \ha \frac{\partial^2 C}{\partial r^2}\mid_{R_h}$.
Since the ball is collapsing, $\nu >0$.

Substituting the above expansions in the expression for $\a '(u)$,
we obtain, to order $(\t _h-\t )^2$,
\be  \label{4-17}
\frac{dU}{du} = a(R_0 - R_h + \t _h - U) +
b(R_0 - R_h + \t _h - U)^2
\te
where
\bea
a &=& (\nu + 1)\kappa \\
b &=& \frac{\kappa}{\nu}\{(3+\nu)\b + (1+\nu)
\frac{\gamma \nu^2}{2\kappa} - \ha A\kappa (1-\nu^2)(1+\nu)\}.
\tea
Note that, for a slowly collapsing ball, $\nu \ll 1$, and hence $a$
reduces to the surface gravity $\kappa$.

Also, to order $(\t _h-\t )$,
\be \label{4-20}
\frac{dv}{dV} = c+d(\t _h + R_h - R_0 - V)
\te
where
\bea
c &=& \frac{A(1+\nu)}{2\nu} \\
d &=& \frac{A}{\nu^2}(\b -\frac{A\kappa}{4}(1-\nu^2)(1+\nu)).
\tea
We consider a regime in which $(\t _h - \t )d \ll c$ so that we may
ignore the second term in \eqn{4-20}. Then we can integrate this
equation to give
\be
v(V) \equiv \b (V) = c_1 + cV
\te
where $c_1$ is an integration constant.

Similarly to lowest order in $\frac{b}{a^2}$ (which turns out to be
the appropriate dimensionless parameter describing deviations from
exact exponential scaling or exact thermal behavior), we integrate
equation \eqn{4-17} to give
\be
U(u) \equiv \a (u) = R_0 - R_h + \t _h + a^{-1} e^{-a(u-c_2)}(1 +
\frac{b}{a^2} e^{-a(u-c_2)}),
\te
$c_2$ being another integration constant.   \\

We are now in a position to obtain explicitly the transformation
between barred and unbarred coordinates, to lowest order in
$\frac{b}{a^2}$. Thus we have
\bea
\overline u &=& \b [\a (u) - 2R_0]  \nn \\
&=& M_1 + M_2 e^{-a(u-c_2)}(1 + \frac{b}{a^2}e^{-a(u-c_2)})
\tea
where
\bea
M_1 &=& c_1 - c(R_0 + R_h - \t _h)  \\
M_2 &=& \frac{c}{a}.
\tea
At the position $r_0^{\ast}$ of the detector, $u=t-r_0^{\ast}$.
Therefore,
defining $\Delta = u'-u$ and $\Sigma = \ha (u'+u) + r_0^{\ast}$,
we may perform
the above transformation to obtain
\be
\overline u' - \overline u = -2M_2e^{ac_2}
\{e^{-a(\Sigma - r_0^{\ast})}\sinh
\frac{a\Delta}{2} + \frac{b}{a^2}e^{-2a(\Sigma - r_0^{\ast} -
\frac{c_2}{2})} \sinh a\Delta\}.
\te
Invoking the identities \eqn{2-10} and \eqn{2-11}, the
function $\zeta_{uu}$ can now be
simplified to yield the near-thermal form
\be
\zeta_{uu} = \frac{e^2}{8\pi}\int_0^{\infty} \frac{dk}{k}
\{\coth(\frac{\pi k}{a})\cos k\Delta (1+\Gamma) - i\sin k\Delta \}
\te
where
\be
\Gamma = -\frac{2bk}{a^3}e^{a(c_2-\Sigma-r_0^{\ast})}
\tanh^2(\frac{\pi k}{a}) \tan k\Delta \sinh a\Delta .
\te
The function $\Gamma$ vanishes at late times
($\Sigma\rightarrow\infty$). Thus
the exact thermal spectrum is recovered at the Hawking temperature
redshifted by the velocity of the surface of the ball, on a time
scale defined by the surface gravity $a$.

\newpage
\section{Discussion}

We now summarize our findings and discuss their implications.
There are four main points made or illustrated here:\\

1) This paper gives {\it a stochastic theoretical derivation of
particle creation},
in the class of spacetimes which possess an event horizon in some
limit.
This approach generalizes the established methods of quantum field
theory  and thermal field theory (in curved spacetimes) to
statistical and stochastic field theory. The exact thermal
radiance cases arising from an exact exponential scale
transformation such as is found in  a uniformly-accelerated detector,
the Schwarzschild black hole and the de Sitter universe,
have been treated in the stochastic theoretical method before
\cite{HM2}.
Here we give the treatment of the moving mirror and the collapsing
mass as further examples. (Thermal radiation in certain classes of
cosmological spacetimes \cite{cosrad} 
inflationary universe will be studied in a following paper.)\\

2) We have shown that in all the examples considered in this class of
spacetimes,
 i.e., accelerated observers, moving mirrors and collapsing masses
(black holes),
those which yield {\it a thermal spectrum of created particles all
involve an
exponential scale transformation}. Thermal radiance observed in one
vacuum arises from the exponential scaling of the quantum
fluctuations (noise)
in another vacuum. This  view espoused by one of us
\cite{HuEdmonton,cgea,jr,Dalian} is illustrated in the examples
treated here.\\

3) The main point of this paper is to show how one can calculate
particle creation in the near-exponential cases, yielding
near-thermal spectra.
These cases are not so easy to formulate conceptually using the
traditional
methods: the geometric picture in terms of the properties of the
event horizons as global geometric entities works well for
equilibrium thermodynamics (actually thermostatics) conditions,
so does thermal field theory
which assumes {\it a priori} a finite temperature condition
(e.g., periodic boundary condition on the imaginary time).
They cannot be generalized to non-equilibrium dynamical conditions
so easily.
In the stochastic theory approach we used, the starting point is
the vacuum fluctuations of quantum fields subjected to
kinematical or dynamical excitations. There is no explicit use
of the global geometric properties of spacetimes:
the event horizons are generated kinematically
by exponential scaling. (Thus, for example, this method can describe
the situations where a detector is accelerated only for a short
duration, whereas one cannot easily describe in geometric terms the
scenario of an event horizon appearing and disappearing.) There is
also no {\it a priori} assumption of equilibrium conditions:
the concept of temperature is neither viable nor necessary,
as is expected  in all non-equilibrium conditions.
Thermal or near-thermal radiance is a result of some specific
conditions acting on the vacuum fluctuations in the system.\\

4) We restrict our attention in this paper to near-thermal conditions
because of technical rather than conceptual limitations.  In the
near-thermal cases
treated here, 
we want to add that the stochastic theoretical method is not the
only way to derive these results. One can alternatively approach with
the global geometric
or thermal field methods, say, by working with generalized
definitions of event horizons or quasi-periodic Green's functions.
However, we find it logically more convincing and technically more
rigorous
to use the stochastic method to define and analyze statistical
concepts like fluctuations and dissipation, correlation and coherence.
Certainly in the fully dynamical and non-equilibrium
conditions, such as will be encountered in the full backreaction
problem
(not just confined to the linear response regime) this method is,
in our opinion, more advantageous than the existing ones.
Even though the technical problems will likely be grave,
(because of the built-in balance between dissipation and fluctuations,
as demanded by a self-consistent treatment), there are no
intrinsic conceptual pitfalls or shortcomings.
These issues and problems are currently under investigation.\\

\noindent {\bf Acknowledgement}
This work is supported in part by the  U S National Science
Foundation
under grant PHY94-21849.
BLH acknowledges support from the General Research Board of the
Graduate School of the University of Maryland and the Dyson Visiting
Professor
Fund at the  Institute for Advanced Study, Princeton.
Part of this work was done while he visited the
Newton Institute for Mathematical Sciences at the University of
Cambridge
during the Geometry and Gravity program in Spring 1994. BLH and AR
enjoyed
the hospitality of the physics department of the Hong Kong
University of
Science and Technology in Spring 1995 when this work was completed.
DK thanks the Australian Vice Chancellors' Committee for its financial
support.

\end{document}